\documentclass[runningheads]{llncs}
\usepackage[margin=1.1in]{geometry}
\usepackage{multirow} 
\usepackage{amsmath}
\usepackage[T1]{fontenc}

\usepackage{graphicx}
\usepackage{hyperref}
\usepackage{color}

\begin{document}
\title{eDOC: Explainable Decoding Out-of-domain Cell Types with Evidential Learning}
\titlerunning{Explainable Decoding Out-of-domain Cell Types}
%
\author{Chaochen Wu\inst{1} \and
Meiyun Zuo\inst{1} \and
Lei Xie\inst{2,3}}
\authorrunning{C. Wu et al.}
%
\institute{Renmin University of China, Beijing, China \and
Hunter College, The City University of New York, New York, NY 10065, USA \and
Weill Cornel Medicine, Cornell University, New York, NY 10065\\
\email{lxie@iscb.org} }
\maketitle              
\begin{abstract}
Single-cell RNA-seq (scRNA-seq) technology is a powerful tool for unraveling the complexity of biological systems. One of essential and fundamental tasks in scRNA-seq data analysis is Cell Type Annotation (CTA). In spite of tremendous efforts in developing machine learning methods for this problem, several challenges remains. They include identifying Out-of-Domain (OOD) cell types, quantifying the uncertainty of unseen cell type annotations, and determining interpretable cell type-specific gene drivers for an OOD case. OOD cell types are often associated with therapeutic responses and disease origins, making them critical for precision medicine and early disease diagnosis. Additionally, scRNA-seq data contains tens thousands of gene expressions. Pinpointing gene drivers underlying CTA can provide deep insight into gene regulatory mechanisms and serve as disease biomarkers. In this study, we develop a new method, eDOC, to address aforementioned challenges. eDOC leverages a transformer architecture with evidential learning to annotate In-Domain (IND) and OOD cell types as well as to highlight genes that contribute both IND cells and OOD cells in a single cell resolution. Rigorous experiments demonstrate that eDOC significantly improves the efficiency and effectiveness of OOD cell type and gene driver identification compared to other state-of-the-art methods. Our findings suggest that eDOC may provide new insights into single-cell biology.

\end{abstract}
\newpage
\section{Introduction}
Advances in single-cell RNA sequencing (scRNA-seq) have facilitated elucidating biological complexity of tissues and organisms. The scRNA-seq provides deep insights into cell-level heterogenities underlying various diseases, like cancer, Alzheimer's disease, and autoimmune disease.
One of critical tasks in scRNA-seq data analysis is to annotate cell types at a single cell resolution. Tremendous efforts in machine learning have been devoted to addressing this problem. However, several challenges remain. First, in a real-world application, some cells may not belong to any known (labeled) cell types, and are called unknown or unidentified cells.  Reliable recognition of these unknown cells may facilitate discovering novel biological processes \cite{papalexi2018single}, and severing as biomarkers for precision medicine and disease diagnosis \cite{kim2021single}. Second, it is important to quantify the reliability or uncertainty of the cell type annotation for risk-sensitive medical applications. Finally, to understand how the novel cell type rise, and explain the prediction, it is needed to interpret the new cell type by highlighting marker genes. Therefore, developing methods that can reliably distinguish known and unknown cells while determining interpretable gene drivers of cell types will harness single cell techniques for advancing basic biology and translational medicine.

\begin{figure*}[!ht]
  \centering
  \includegraphics[width=\textwidth]{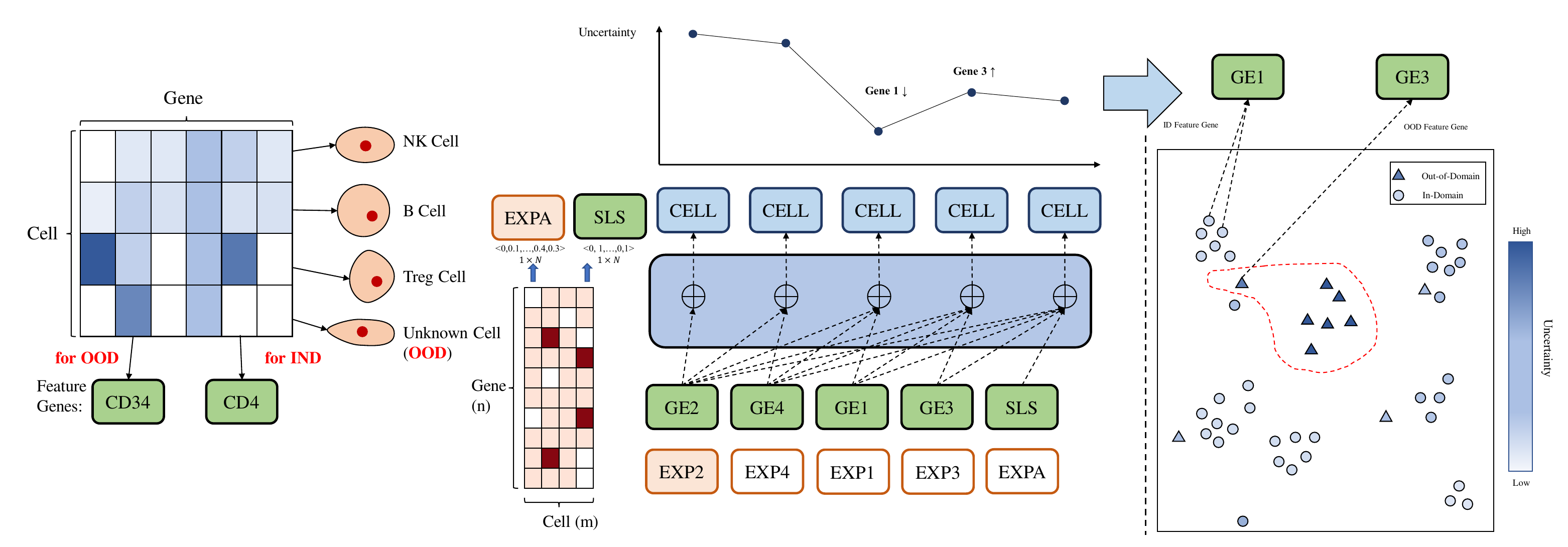}
  \caption{eDOC architecture and illustration of evidential learning to find OOD cells and detect marker genes in OOD cells. If Gene 1 causes Uncertainty drop significantly, so Gene 1 is one of marker genes for IND cell types. Adding Gene 3 causes the increase of uncertainty so Gene 3 is one of marker genes for OOD cells. Whether a cell is OOD or IND cell depending the last position in the transformer when all genes are visible to the model.}
  \label{fig-1}
\end{figure*}

In this studies, we treat the unknown cell type detection as an out-of-domain (OOD) sample detection task similar to OOD detection for languages \cite{dai2007co} or images \cite{li2021semantic}. Although various types of OOD detection methods exist, quantifying the uncertainty of cell type annotation for a new cell and finding which features make a sample different from known cell types as an OOD case remain challenging. For In-Domain samples,  interpretability techniques, like SHAP, can assess feature importance of a case given a trained model. However, these methods are not designed for unseen OOD samples that are significantly different from training dataset. For example, in a training set with $K$ classification labels for all sentences, and we can use current interpretability methods to show which words contribute to $K$ label classifications. However, when we meet the sentence belongs to an unknown label (OOD case) in the test set, we cannot tell which words contribute it be a OOD sample. For the unknown cell type annotation, task are summarized as: recognize if a cell is a unknown cell type, and identify which genes characterize a cell to be a new cell type. The left part of Fig. \ref{fig-1} illustrates the task for scRNA-seq data analysis.

Currently, transformer-based models are successful in not only various language tasks but also analyzing scRNA-seq data\cite{yang2022scbert,gong2024xtrimogene,cui2024scgpt}. It is important to note a key difference between language and gene expression data: word order is crucial for text processing \cite{wang2020encoding} and benefits self-supervised pre-training, but there is no inherent order for genes in scRNA-seq data. Therefore, we leverage this characteristic by employing the attention mask mechanism of Transformer \cite{vaswani2017attention} to test the effects of individual genes being present or absent. This approach allows us to efficiently explain cell types based on gene expressions.

Based on above rationale, we introduce a new scRNA-seq analysis method called \underline{\textbf{e}}xplainable \underline{\textbf{D}}ecoder for \underline{\textbf{O}}ut-of-domian \underline{\textbf{C}}ell (eDOC). We use the Transformer architecture as a backbone to encode a single cell RNA expression data; instead using the $[CLS]$ token for cell-type annotations, we train and inference cell types as a decoder only language model and output cell types for all token positions. We adopt evidential learning to train our model, which enables us to observe how individual genes contribute to both IND cell types and the OOD cell type with uncertainty change. In practice, our method does not require researchers to definitively identify all cell types with existing samples. They can use eDOC to classify uncertain or rare cells as OOD samples and decipher gene drivers associated with the new cell type. Advantages of our model are summarized as below:
\begin{itemize}
    \item Our method have a superior performance for OOD cell types detection without supervised training with OOD samples. Additionally, it can also accurately annotate known cell types.
 
    \item To our best knowledge, our method can interpret OOD cell types with marker genes in a single-cell resolution for the first time, and it can also interpret IND cell types.

    \item Our method is fast, simple and easy-to-use, and it only need transcriptomics data for training and inference without the requirement of additional knowledge or pre-training.
\end{itemize}

\section{Related Work}
\subsection{Cell Type Annotation for scRNA-seq}
There are primarily three categories of methods for the cell type annotation. The first type uses marker genes to specify cell types. For example, CD4 is one of the marker genes for T lymphocytes (T cells). This method relies on researchers' prior knowledge and databases, which is laborious and prone to error and bias \cite{huang2021evaluation}. The second type is the cluster-then-annotate approach, which is the most popular one. This method uses similarities of cells' gene expression profiles and assigns type labels to all cells within each cluster using genes' averaged expression levels. However, this approach is limited by low-coverage sequencing data from experiments \cite{hou2019scmatch}. The third type is supervised classification, and some of them are based on deep learning. scBERT\cite{yang2022scbert} uses bidirectional encoder that is pre-trained by large-scale scRNA-seq data to classify cell types. scGPT \cite{cui2024scgpt} use the decoder architecture like GPT to pre-train model and annotate cell types using the gene expression of a cell. Several transformer-based methods highlight their interpretability with the attention score \cite{yang2022scbert,chen2023transformer}, but whether attention scores can be used for interpretation \cite{jain2019attention,wiegreffe2019attention} is controversial. Genes that are important by rarely expressed may be masked by other genes and have low attention scores.

\subsection{Out-of-domain Detection}
Various strategies for OOD detection contribute to machine learning model deployment in the real world. Maximum softmax probabilities(MSP) \cite{hendrycks2022baseline} is a baseline that uses MSP score to distinguish between IND and OOD samples; LMCL \cite{lin2019deep} apply margin loss to learn discriminative deep features. Supervised Contrastive Learning (SCL) \cite{zeng2021modeling} tried to pull different samples with the same label together, and it additionally improved K-Nearest Neighbors (KNN) based  \cite{zhou2022knn,sun2022out} and  Mahalanobis distance-based \cite{sehwag2021ssd} OOD detection. Besides the above methods, Uncertainty learning is a promising solution to OOD detection \cite{charpentier2020posterior,ulmer2021know}. Some cell type annotation methods mentioned their ability in unknown cell detection \cite{yang2022scbert,chen2023transformer,jia2023scdeepinsight}, and their methods set a cut-off for the probability score of the output layer, which is similar to MSP. 

\section{Method}
\subsection{Problem Definition}
We first consider the IND cell type annotation as a supervised multi-class classification task; the input and output are defined as $x_i$ and $y_i \in \{1,2, \dots K\}$, respectively. Here, $x_i$ is a vector of the expression of $N$ genes, and $K$ is the number of cell types. An IND training set is $D_{IND}=\{(x_i,y_i)\}_{i=1}^M$, and each sample $(x_i,y_i)$ is drawn from IND distribution $P_{IND}$. The goal of our method is to use $D_{IND}$ as the training set and classify $K$ known cell types if $x_i$ is drawn from $P_{IND}$, and it can detect unknown/novel cells ($x_i$) from a different distribution $P_{OOD}$ and $y_i \not\in \{1,2, \dots K\}$. The input of scRNA-seq expression data for the training is a $N \times M$ matrix, which contains $M$ cells and $N$ genes for each cell.   

For explaining a cell with $\hat{N}$ expressed genes, we use decoder to output $\hat{N}+1$ uncertainty score $\{u\}_{n=1}^{\hat{N}+1}$ for each tokens. Explaining $n$th gene, except the first one, can be simple done by computing $u_n-u_{n-1}$. The overall architecture is summarized in Fig. \ref{fig-1}.

\subsection{Gene Embedding}
Similar to word sentences, we use the embedding layer to embed $\hat{N}$ genes to $\{g_n\}_{n=1}^{\hat{N}}$ and use fully-connected to encoding gene expression to $\{h_n\}_{n=1}^{\hat{N}}$. We append the $SLS$ token behind the last position of a gene sequence as the $(\hat{N}+1)$th element, which is the compacted representation of $N$ genes and their expression. 
\begin{equation}
    g_{SLS} = FC(H_{OneHot}^N), H_{OneHot}^N \in R^{1 \times N} 
\end{equation}
\begin{equation}
    h_{SLS} = FC(H^N), H^N \in R^{1 \times N} 
\end{equation}
$H^N$ is one cell's gene expression data that is directly collected from the $N \times M$ matrix, and $H_{OneHot}^N$ is the One-Hot version of $H^N$, that represents whether a gene appear (value $1$) in $\hat{N}$  or not (value $0$). We concatenate $g_n$ and $h_n$ to generate fused embedding $G_{\hat{N}+1}\in R^{(\hat{N}+1)\times d_{g}}$, and $d_{g}$ is the dimension of gene embeddings:
\begin{equation}
    G_{\hat{N}+1} = FC([\{g_n\}_{n=1}^{\hat{N}+1} \cdot \{h_n\}_{n=1}^{\hat{N}+1}])
\end{equation}

\subsection{Transformer Decoder}
We used the decoder-only structure from Transformer \cite{vaswani2017attention} to encode gene and expression embedding:
\begin{equation}
    \Theta = Transformer(G_{\hat{N}+1})
\end{equation}

Then, a fully connected layer is applied to the Transformer output ($\Theta$) to produce $E$ as the model output.
\begin{equation}
    E = FC(\Theta), E \in R^{K \times (\hat{N} + 1)}
\end{equation}

Because of the presence of the attention mask, the decoder model is constructed by $N+1$ cell type annotators:
\begin{equation}
    y_n = CTA^{n}(g_1 \cdot H_1,\cdots,g_n \cdot H_n)
\end{equation}

In the $n$th position, the model can only use the first $n$ genes in the input sequence for estimating cell types, so eDOC is constructed by a strong cell type annotator in $\hat{N}+1$ position and $\hat{N}$ ``weak" annotators from $1$ to $\hat{N}$ positions.  

\subsection{Model Training and Evidential Learning for Uncertainty Score}
We employ the evidential deep learning (EDL) \cite{sensoy2018evidential} to train the model and compute the uncertainty score for each output from the decoder. For the $n$th decoder output $E_n=[e_1,\cdots,e_K]$, we got Dirichlet strength $S_n$ by:
\begin{equation}
    S_n =\sum_{k=1}^{K}(e_k + 1)
\end{equation}
The uncertainty score can be computed by:
\begin{equation}
     u_n+\sum_{k=1}^{K}\frac{e_k}{S_n}=1, u_n = \frac{K}{S_n}
\end{equation}
The EDL loss function is:
\begin{equation}
    \mathcal{L}^{edl}_{n} = \sum_{k=1}^{K}y_k(\log(S_n) - \log(e_k + 1))
\end{equation}
Where $y_k$ is one-hot vector encoding the ground-truth class of cell type. Our method trains $\hat{N}+1$ cell type annotators together with:
\begin{equation}
    \mathcal{L}^{cta} = \sum_{n=1}^{\hat{N}+1}(\mathcal{L}^{edl}_{n})
\end{equation}

\subsection{Cell Type Output and Interpretation}
The last decoder output $E_{\hat{N}+1}$ is applied for producing IND labels, because in this position model can use maximum gene information and 
 their expression to make the best decision. The $u_{\hat{N}+1}$ can be used to label OOD sample:
\begin{equation}
    \begin{cases}
    IND \quad u_{\hat{N}+1} \leq \lambda \\
    OOD \quad u_{\hat{N}+1} > \lambda \\
    \end{cases}
\end{equation}

Where $\lambda$ is a threshold parameter. We use $u_n$ to interpret $n$th gene effect on IND and OOD cell by $u_n^{diff}$:
\begin{equation}
    \begin{cases}
    u_n^{IND}=u_{n-1} - u_n,  \quad  IND\\
      u_n^{OOD}=u_{n} - u_{n-1}, \quad OOD \\
    \end{cases}
\end{equation}

This metric is straightforward: if the $n$th gene has a high $u_n^{IND}$ and $u_{n-1} - u_n > 0$, this gene is important for IND cell types; if the $n$th gene have a high $u_n^{OOD}$ and $u_{n-1} - u_n < 0$, the ‌appearance of this gene causes the confuse of cell type annotations, so it is important for OOD.

\subsection{Order Shuffling for Robust Interpretation}
eDOC can produce the cell type annotation and $\hat{N}-1$ gene interpretation with a single run. Although the absence of order in gene set helps the model show genes effects in a single run, some driver genes, which are assigned front positions, have a strong effect that hides other genes' contribution (details see experiments). To ensure our method can observe gene effects in different positions, we can change the order of the input sequence. The simplest way is to randomly shuffle the order of the gene input sequence and run the model multiple times. Users can also manually move genes they are interested in the front of other genes that are known to be unimportant. In the training phase, we randomly select a maximum of $\hat{N}$ genes as the input sequence, changing the selection of genes and their order for different epochs.

\subsection{Time Complexity and Deployment}
Many scRNA-seq datasets only allow restricted access from authorized people, so researchers have to use limited computational resources to train models and perform analysis. eDOC use limited length ($\hat{N}+1$) of the sequence as the model input, which decrease the computation cost:
\begin{equation}
    \hat{N}+1 \ll N
\end{equation}
By running eDOC that costs $T$ time, eDOC can produce results of IND cell type annotations, OOD detections, and the interpretation of $\hat{N}-1$ genes at the same time. If we want to perform robust interpretation that quantifies how genes contribute to cell types comprehensively, the time complexity of eDOC is linear.  

\section{Experiments}
\subsection{Datasets and Experiment Settings}
For the OOD detection of cells, we take Zheng68K for peripheral blood mononuclear cells (PBMC) \cite{zheng2017massively} as the benchmark. It includes 11 PBMC cell types. We exclude three cell types: CD4+ T Helper2, CD34+, and CD4+ /CD45RA+ /CD25- Naive T as OOD cells from the original dataset. We split eight types of IND cells into three portions for training, validation, and testing. By merging different numbers of testing IND cells with OOD cells, we obtain three test datasets with 25\%,  50\%,  and 75\% percentages of IND cells to evaluate the OOD detection. In the IND annotation setting, we take all cell types from Zheng68K and  Segerstolpe \cite{segerstolpe2016single} for training and testing. We use the F1 score as the metric for OOD detection as previous studies \cite{zeng2021modeling,zhou2022knn}. For testing IND cell type annotations,  we use macro-averaged F1 and precision as metrics. 

The $\hat{N}$ we used for IND and OOD experiments are 512, and selected $\hat{N}$ and their order change with training epochs. The Transformer backbone is constructed by 4 decoder layers with multi-head attention (8 heads). The threshold for OOD ($\lambda$) is set to $0.1$. The dropout rate we use is 0.5. We use a single RTX 3090Ti or L40S GPU to train eDOC.

\subsection{OOD Cell Detection}
We compare eDOC with state-of-the-art OOD detection methods: MSP \cite{hendrycks2022baseline}, LMCL \cite{lin2019deep}, supervised contrasting learning (SCL) with Local Outlier Factor (LOF) and Gaussian Discriminant Analysis (GDA) \cite{zeng2021modeling}, KNN with SCL (KNN+) \cite{sun2022out}, a Transformer-based cell type annotation method scRNA-seq (TOSICA) \cite{chen2023transformer}, scBERT, a meta-learning approach (MARS) \cite{brbic2020mars}, and an approach that use uncertainty quantification (scPoli) \cite{de2023population}. We show results in Table \ref{table-1}
\begin{table}[!ht]
  \caption{Performance Comparison of Out-Of-Domain Detection, The best and second best values are in bold and underscore, respectively}
  \centering
  \begin{tabular}{|c|c|c|c|}
  \hline

    Methods & 25\%(F1) & 50\%(F1) & 75\%(F1)   \\
     
    \hline
    MSP(Transfomrer) & 40.00 & 66.67 & 85.71   \\
    TOSICA & 44.25 & 67.13 & 82.06  \\
    scBERT & 41.79 & 67.06 & 85.12  \\
    scPoli & 42.04 & 68.38 & \underline{86.64}  \\
    MARS & 40.00 & 66.55 & 85.67  \\
    LMCL & 42.42 & 67.47 & 83.92   \\
    SCL+LOF & 43.28 & \underline{68.40} & 84.09   \\
    SCL+GDA & 42.24 & 67.03 &  85.35  \\
    KNN+ & \underline{44.95} & 67.89 &  85.73  \\
    \hline
    eDOC (ours) & \textbf{60.08} & \textbf{76.25} & \textbf{88.56}  \\
    \hline
  \end{tabular}
  
  \label{table-1}
\end{table}
The comparison result for the OOD detection proves the outstanding OOD detection ability of eDOC. When the percentage of IND cells decreases, we observe a significant F1 score decrease for other models, but our method eDOC only has a moderate decrease. When there are only 25\% IND cells, the F1 score of other methods is only from 0.4 to 0.45, but eDOC is around 0.6, a 50\% improvement. Although MSP has been adopted for various unknown cell detection methods, its performance is suboptimal in the presence of a larger number of OOD cells. 

Additionally, we run our model for the OOD cell detection in different scRNA-seq datasets from different tissue samples, including the pancreas, liver, heart, and lung, and our method significantly outperform the baseline methods. We present these results in the Appendix.   

\subsection{IND Cell Type Annotation}
We also compared our method with SOTA IND cell type annotation methods: ACTINN \cite{ma2020actinn}, Scanpy \cite{wolf2018scanpy}, scVI \cite{lopez2018deep}, singleCellNet \cite{tan2019singlecellnet}, xTrimoGene \cite{gong2024xtrimogene}, scBERT \cite{yang2022scbert}, 
  CellTypist \cite{dominguez2022cross}. The result is present in Table \ref{table-2}.
\begin{table*}[!ht]
  \caption{Performance Comparison of In-Domain Cell Type Annotation. The best and second best values are in bold and undersocre, respectively}
  \centering
  \begin{tabular}{ccccccccccc}
  \hline
    \multirow{2}{*}{Model} & \multicolumn{2}{c}{Zheng68K}  & \multicolumn{2}{c}{Segerstolpe}  \\ 
    \cline{2-3} \cline{4-5} 
     & Prec(ID) & F1(ID)  & Prec(ID) & F1(ID) \\
     
    \hline
    
    ACTINN & 67.20 ± 0.21 & 64.86 ± 0.41   & 75.45 ± 0.18 & 72.19 ± 0.73 \\
    Scanpy & 61.11 ± 0.17 & 54.74 ± 0.85   & 62.74 ± 0.00 & 53.98 ± 0.00    \\
    scVI & 48.83 ± 0.05 & 48.43 ± 0.08  & 51.01 ± 0.22 & 52.08 ± 0.16    \\
    singleCellNet & 64.52 ± 0.13 & 59.82 ± 0.27   & 75.51 ± 0.96 & 80.55 ± 0.76    \\
    xTrimoGene & 73.35 ± 2.26 & \textbf{73.54} ± 1.89   & \textbf{81.12} ± 0.09 & \textbf{81.40} ± 0.08 \\
    scBERT & 70.29 ± 1.15 & 66.95 ± 0.77   & 68.18 ± 7.36 & 67.03 ± 6.53 \\
    CellTypist & \underline{74.54} ± 0.09 & \underline{71.51} ± 0.38   & 79.23 ± 0.03 & \underline{81.17} ± 0.01    \\
    \hline
    Ours & \textbf{76.95} ± 2.55 & 70.67 ± 0.35  & \underline{80.88} ± 0.92 & 80.57 ± 0.46    \\
    \hline
  \end{tabular}
  
  \label{table-2}
\end{table*}

When comparing our method with SOTA methods for the IND cell type annotation, the performance of our method remains outstanding. For the Zheng68K dataset, eDOC got the best result in precision score. For the Segerstolpe dataset, eDOC ranks in the second, but performance difference from the best xTrimoGene is not significant (p-value = 0.29). For F1 scores evaluated by Zheng68K and  Segerstolpe dataset, its result is comparable to the best SOTA.

\subsection{How does eDOC Explain and Label}
\begin{figure}[!ht]
  \centering
  \includegraphics[width=\textwidth]{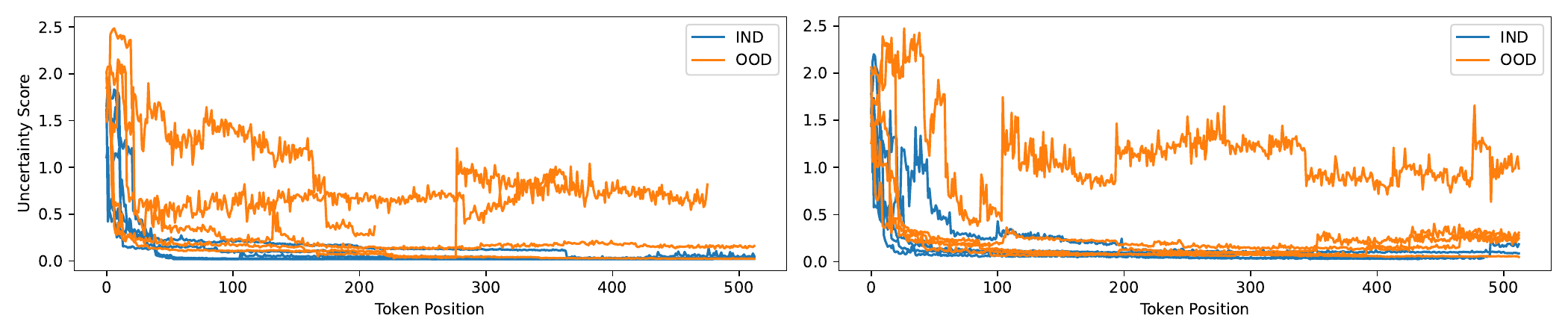}
  \caption{Two examples of uncertainty score changes. In each plot, we randomly select 5 OOD cells and 5 IND cells to draw uncertainty score $u_n$ changes with $n$ genes. Each line represent a cell. }
  \label{fig-2}
\end{figure}

Our method provides an intuitive insight into how gene expression contributes to cell type annotations. By removing masks of attention, the decoder can use more genes to decide cell types. For IND cells, more genes make the model more ``confident" in its cell types. Conversely, some genes make the model feel ``confused" for OOD cells. When we use $u_n$ to describe this difference, we can observe $u_n-u_{n-1} < 0$ or $u_n-u_{n-1} > 0$ if the $n$th gene makes model more ``confident"  or more ``confused", respectively. We use the last position $\hat{N}+1$ as the cell representation because, in this position, the model can view all gene expressions, making it the most informative cell type annotator. We use eDOC to produce $u_n$ for each token position to compare OOD cells and IND cells in Fig. \ref{fig-2}.

With the increased number of genes involved in cell type annotations, we can observe a decrease in $u_n$. For IND cells, the $u_n$ finally drops to a low value. For OOD cells, some genes make $u_n$ increased significantly, and the model finally got a high $u_{\hat{N}+1}$. This obvious difference helps us explore genes for IND and OOD cells and distinguish between IND and OOD cells.

\begin{figure*}[!ht]
  \centering
  \includegraphics[width=0.95\textwidth]{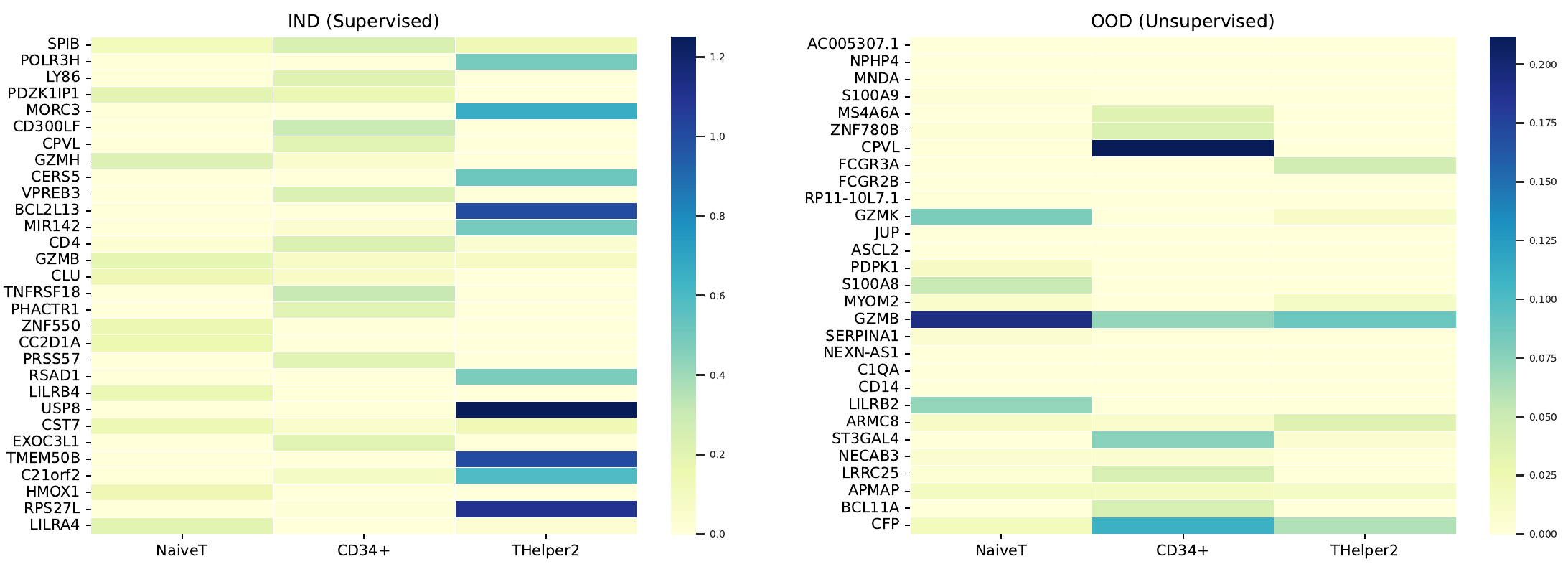}
  \caption{Illustration of genes that are highlighted by eDOC for IND (left) and OOD (right) experiments. }
  \label{fig-3}
\end{figure*}

In  Fig. \ref{fig-3}, we highlight genes with outstanding $u_n^{IND}$ for IND cell type annotation and OOD detection tasks. For the IND cell type annotation, we compute average $u_n^{IND}$ for each cell type and pick the top 10 genes with the highest $u_n^{IND}$ for three types of cells to draw the heatmap in the left part of Fig. \ref{fig-3}. For the OOD detection task, after we get the $u_n^{OOD}$ for all cells, we filter OOD cells from the test data and split OOD cells by their real cell labels, and we compute the average $u_n^{OOD}$ for each OOD cell to draw the heatmap with top 10 genes in the right part of Fig. \ref{fig-3}.

In the heatmap, genes show obviously different $u_n^{IND}$ and $u_n^{OOD}$ patterns among cell types. Some genes appear in both unsupervised OOD detection and supervised IND cell type detection. The functional roles of these top-ranked genes in their corresponding cell types are supported by published works. GZMB is highlighted by both IND and OOD study. It is a marker gene for T cell and a potential therapeutic target for the treatment of rheumatoid arthritis \cite{zhang2023exploring}. CPVL involves in different diseases progression like glioma \cite{yang2021cpvl}. Although supervised IND cell type annotation and unsupervised OOD detection produce a different gene set, some genes from two gene sets are involved in the same biological processes. For example, BCL2L13 from IND and BCL11A from OOD encode proteins for apoptosis \cite{handschuh2021transcript}. 

\subsection{Robust Interpretation with Single Cell Cases}
eDOC allows us to investigate how different genes contribute to IND cell type annotation and OOD cell detection robustly by simply changing the input order of genes. Because cell types are decided by multiple genes, and some genes may hide other genes' effects. eDOC can improve the gene interpretability for cell type annotation by placing genes in early $n$ positions, and genes located in late positions (like $n+1$ and $n+2$) make no contributions to cell type annotations in the $n$th position because the attention mask blocks the effect of genes in the late position. This allows us to reveal how genes affect cell types without being affected by non-marker genes. Many deep learning-based methods use attention weights for models' interpretability.  Here, we  compare the explanation from eDOC with that by attention weights. For IND cell type annotation, we select three cell types, and each cell type has five expressed important marker genes for this cell type, which are supported by literature\footnote{References in the Appendix.}. Fig. \ref{fig-4} shows that difference between eDOC and attention weights.

\begin{figure*}[!ht]
  \centering
  \includegraphics[width=\textwidth]{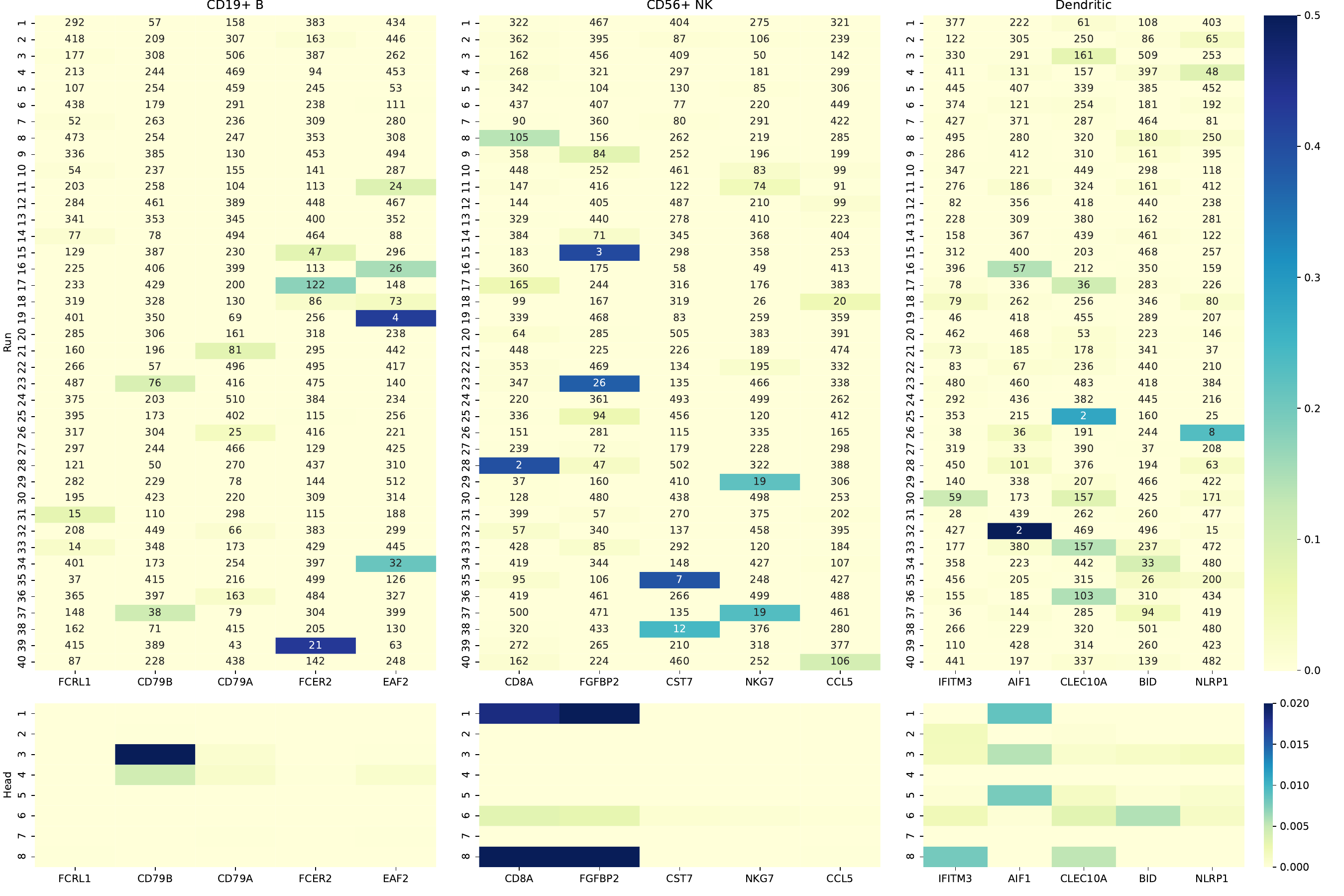}
  \caption{Examples of identified marker genes by eDOC (top) and attention weights (bottom) for three different IND cells. Each heatmap panel represents a cell with five marker genes for the cell. Heatmaps in the top row show eDOC's $u_n^{IND}$ with 40 runs, and the number in each heatmap panel represents the $n$th position. Heatmaps in the bottom row show attention weights for all heads (8) of the Transformer model. }
  \label{fig-4}
\end{figure*}

\begin{figure*}[!ht]
  \centering
  \includegraphics[width=\textwidth]{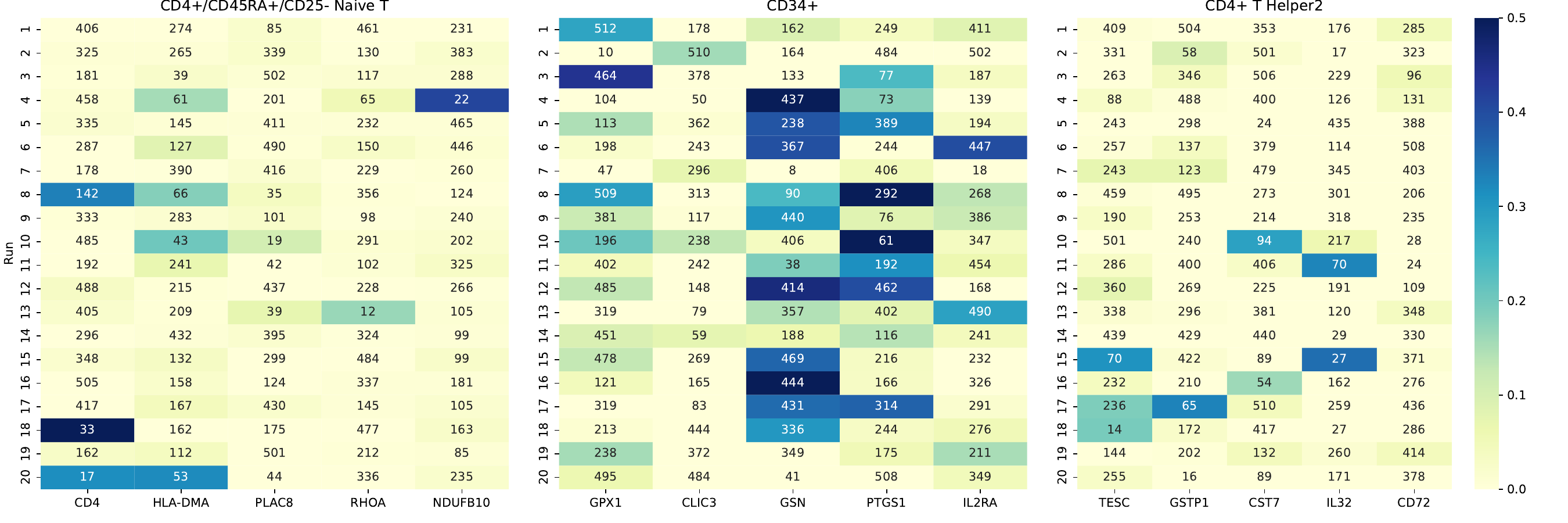}
  \caption{Examples of identified marker genes with eDOC for three different OOD cells. Each heatmap panel represents a cell with five marker genes for the cell. The heapmap shows eDOC's $u_n^{OOD}$ with 20 runs, and the number in each heatmap's cell represents the $n$th position.}
  \label{fig-5}
\end{figure*}

From heatmaps, attention weights only focus on a portion of gene sets but miss a number of important genes. It may be caused by some genes with high attention weights hiding other genes' contributions. Instead eDOC performs a robust explanation for marker genes by running multiple times with different orders for gene inputs. For the explanation of the NK cell, both attention weights and eDOC show the importance of CD8A and FGFBP, but CST7, NKG7, and CCL5 are overlooked by attention weights. The heatmap that is produced by eDOC suggests when CD8A and FGFBP are in front of other genes, they can mask the importance of other genes. eDOC eventually highlights the importance of CST7, NKG7, and CCL5 by putting them in front of CD8A and FGFBP, but attention weight misses these genes. Thus, eDOC can use $u_n^{OOD}$ to identify marker genes for OOD cells, which would be missed by attention weights. In addition, eDOC can reliably identify marker genes for IND cell types.  Fig \ref{fig-5} shows that five marker genes identified for three OOD cell types by eDOC's $u_n^{OOD}$.

\section{Conclusion}
In this paper, we introduce eDOC, a new method that can robustly detect novel cell types and identify marker genes simultaneously. eDOC not only outperforms state-of-the-art methods in cell type annotations but also for the first time can highlight important genes in an OOD setting.  eDOC will facilitate biomedical researchers gaining new insights into disease mechanisms from expensive but invaluable sequencing projects. Our work shows that progress in NLP and deep learning can benefit biomedical research. New AI methods have potential to address OOD challenges and improve the interpretability of biomedical data analysis.

eDOC provides a new strategy for analyzing scRNA-seq data by treating uncertain cells as out-of-distribution (OOD). This allows us to apply eDOC to explore novel and rarely appearing cell types and characterize them with marker genes. In the future, we will conduct experiments to examine how the marker genes discovered by eDOC contribute to cell phenotypes and the roles of identified cells and genes in diseases and treatments.
%
%
%
\bibliographystyle{splncs04}
\bibliography{main}

\end{document}